# Investigating laser induced phase engineering in $MoS_2$ transistors

Nikos Papadopoulos, Joshua O. Island, Herre S. J. van der Zant and Gary A. Steele

*Abstract*—**Phase engineering of $MoS_2$ transistors has recently been demonstrated and has led to record low contact resistances. The phase patterning of $MoS_2$ flakes with laser radiation has also been realized via spectroscopic methods, which invites the potential of controlling the metallic and semiconducting phases of $MoS_2$ transistors by simple light exposure. Nevertheless, the fabrication and demonstration of laser patterned $MoS_2$ devices starting from the metallic polymorph has not been demonstrated yet. Here, we study the effects of laser radiation on $1T/1T'$-$MoS_2$ transistors with the prospect of driving an *in-situ* phase transition to the $2H$-polymorph through light exposure. We find that although the Raman peaks of $2H$-$MoS_2$ become more prominent and the ones from the $1T/1T'$ phase fade after the laser exposure, the semiconducting properties of the laser patterned devices are not fully restored and the laser treatment ultimately leads to degradation of the transport channel.**

## I. Introduction

The transition metal dichalcogenides (TMDCs) form a large family of layered materials that have been studied extensively in the last few years [1],[2],[3],[4]. $2H$-$MoS_2$ is one of the most well-known in this family with a direct optical bandgap of 1.8 eV [5], which becomes indirect and decreases as the number of the layers increases, reaching 1.3 eV in bulk [6]. These properties render $2H$-$MoS_2$ ideal for applications such as field-effect transistors (FETs) [7], photodetectors [8, p.], and light emitting diodes (LEDs) [9]. A striking difference between TMDCs and other 2D materials like graphene is the polymorphism of these materials [2]. Naturally occurring semiconducting $MoS_2$ has a trigonal prismatic structure ($2H$-$MoS_2$). Another known polymorph is $1T$-$MoS_2$ with an octahedral geometry [10],[11], which has metallic properties and stabilizes with lattice distortion by forming the so-called $1T'$-$MoS_2$, where clustering of Mo atoms takes place with the formation of various superstructures [12]. A known route to obtain the $1T/1T'$ phase is via chemical doping usually by using *n*-butyl lithium (BuLi) were Li atoms donate an electron to the Mo atoms [13]. The dynamics and the mechanisms of intercalation and the phase transformation in $MoS_2$ has been studied by several groups in recent years [12],[14],[15],[16],[17]. These studies have shown that the $1T$ and $1T'$ phases coexist ($1T/1T'$ phase) and the $1T/1T'$ phase is present even after removal of the lithium with more metallic properties than the natural polytype [13].

$1T/1T'$-$MoS_2$ is metastable with a relaxation energy of ~1 eV and relaxes to the $2H$-phase with annealing above 95 $^o$C [11] or with extensive aging [18]. Another route to induce a metallic to semiconducting transition is via laser heating as shown recently by two different groups [19],[20]. This approach is intriguing as the phase transformation takes place locally and provides the opportunity to form in-plane heterostructures [21]. This type of heterostructure is interesting for TMDCs as $1T/1T'$ contacted $MoS_2$ [22], [23] and $WSe_2$ [24] FETs show lower contact resistance and superior device characteristics [25]. Also, such heterostructures could be used for catalytic processes. Nevertheless, the fabrication of such planar heterostructures via laser patterning and their properties have not yet been explored.

This work was supported in part by the Netherlands Organisation for Scientific Research (NWO) and the (Ministry of Education, Culture, and Science (OCW). (Corresponding author: Nikos Papadopoulos)

N. Papadopoulos, J. O. Island, H. S. J. van der Zant and G. A. Steele are with the Kavli Institute of Nanoscience, Delft University of Technology, Lorentzweg 1, Delft 2628 CJ, The Netherlands (e-mail: n.papadopoulos@tudelft.nl).

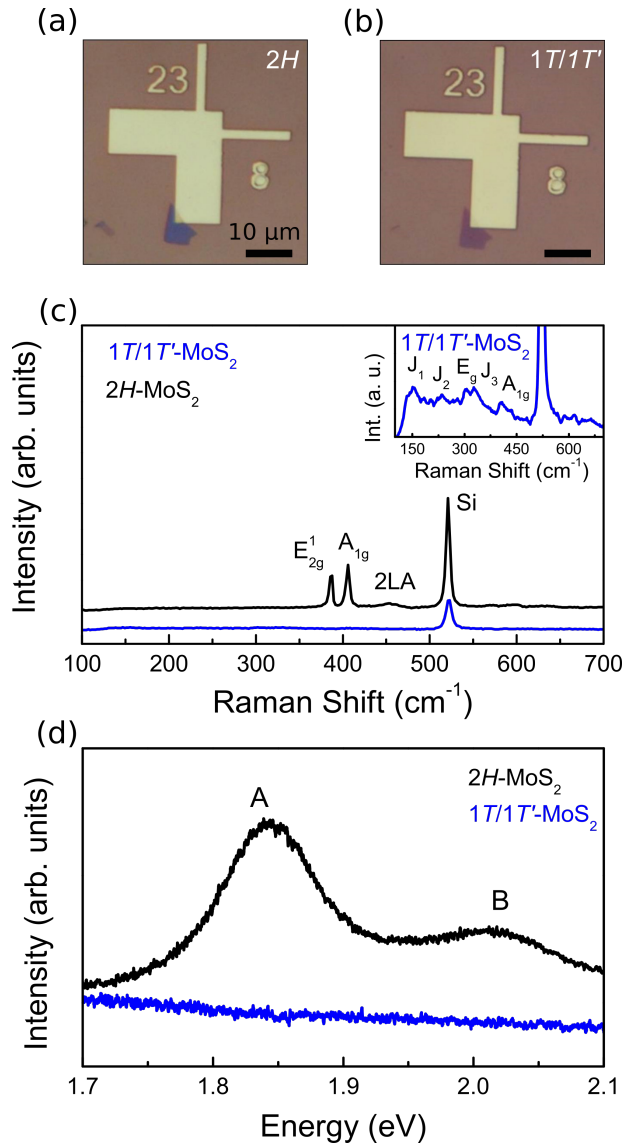

Fig. 1. Semiconducting to metallic transition in $MoS_2$ via chemical doping. Optical images of a few-layer $MoS_2$ before (a) and after (b) the treatment with *n*-butyl lithium where the change in the color is attributed to the phase transition. (c) Raman spectra of the $2H$ and $1T/1T'$ phases of $MoS_2$. The inset shows the spectra of the $1T/1T'$-$MoS_2$ from with a different range on the y-axis. The $J_1$, $J_2$ and $J_3$ peaks from the $1T/1T'$ phase are visible. (d) Photoluminescence spectra of the $2H$ and $1T/1T'$ of $MoS_2$. The A and B excitonic peaks in the case of the $1T/1T'$-$MoS_2$ are quenched.

Here, we investigate patterning of the semiconducting phase of $MoS_2$ via laser induced local heating in a few layer FET. We use green laser light at 515 nm and we monitor the transformation processes via Raman spectroscopy. Moreover, we extract the range of the laser power that should be used for such process by analyzing the Raman spectra. Thereafter we investigate the laser scribed planar $MoS_2$ heterostructures in various FET devices and the impact of the consecutive phase transitions on the characteristics of the devices.

## II. Results and discussion

The semiconducting to metallic transition in $MoS_2$ was achieved by chemical doping in *n*-butyl lithium solution (1.6 M in hexane, Sigma Aldrich) inside a glove box environment. Usually though with this treatment the samples can reach 50% of the $1T/1T'$ phase within a flake [21],[26]. We used optical imaging, Raman, photoluminescence spectroscopy as well as electrical measurements for the characterization of the transition. In Fig. 1(a) and (b) the optical images of a thin $MoS_2$ flake (4-5 layers) before and after the immersion in BuLi show the change in the optical properties of the material. Unambiguously the color and the contrast of the flake after the bath has changed similar to other studies [27]. Raman spectroscopy is a powerful technique to study structural changes and chemical bonds in materials and molecules. The different crystal structure that arises after the chemical doping can be easily probed with Raman spectroscopy. In Fig. 1(c) the Raman spectra of $2H$-$MoS_2$ and $1T/1T'$-$MoS_2$ are depicted, which were obtained with low power (<0.05 $mW/\mu m^2$) to avoid any heating effects. In the case of the $1T/1T'$-$MoS_2$, the absence of the in-plane $E_{2g}^1$ peak and the weak $A_{1g}$ peak at 404 $cm^{-1}$ together with the presence of small features at 155, 233, 331 $cm^{-1}$, confirm the structural arrangement of the atoms in an octahedral and a distorted octahedral lattice (see inset of Fig. 1(c)). The three weak peaks are the $J_1$, $J_2$ and $J_3$ modes, which arise from the formation of a $2a_0 \times a_0$ superlattice, probably as a result of a Peierls [28] or a Jahn Teller instability [12] and the clustering of Mo atoms into chains [29],[18],[30]. Apart from the quenching of the Raman peaks of the $2H$ phase, the A and B exciton peaks of the photoluminescence spectra were also quenched (Fig. 1(d)), confirming the partial change to the $1T/1T'$ polymorph.

For the transformation and the recovery of the semiconducting properties of the $MoS_2$ flakes and FETs, we used laser radiation at 515 nm with a 100x objective with a spot of ~0.5 $\mu m \times \mu m$. To find out the right power that can lead to the transformation without inducing layer thinning [31] or damaging [32], we monitored in real-time the Raman spectrum for different laser powers. We kept the exposure time the same (30 sec) in order to avoid any time-dependent effects. Figure 2(a) shows the Raman spectra of a 6 nm thick flake at different powers, which are normalized to the incident laser power. At low power, the $J_1$, $J_2$ and $J_3$ peaks at 155, 230, and 330 cm$^{-1}$ can be clearly seen, while the $E_{2g}^1$ and $A_{1g}$ peaks have low intensity, indicating the presence of a small residual areas of 2$H$-phase. All peaks have higher intensity than the ones in figure 1 due to the larger thickness of the flake in this case. From 0.3 mW and above, the intensity of the $E_{2g}^1$ and $A_{1g}$ peaks from the 2$H$ phase increases, while the intensities of the peaks belonging to the 1$T$/1$T'$ polymorphs decrease. This change indicates the gradual increase (decrease) of the 2$H$ (1$T$/1$T'$) phase content within the flake.

The response of the MoS$_2$ to different laser powers can be better understood by investigating the behavior of the integrated intensity ratio of the $A_{1g}$ to the $J_3$ peak. We choose these two peaks for a more accurate fit at low fluence but the analogous behavior is observed for the ratio of the $E_{2g}^1$ to the $J_1$ or $J_3$ peak. Figure 2(b) shows a logarithmic plot of the intensity ratio of $A_{1g}$ to the $J_3$ peak as a function of the of the laser power for two different flakes with thickness of 6 and 23 nm. As can be seen, the behavior is similar in the two flakes and the ratio of the intensities does not change up to 0.4 mW. Between 0.4 and 1 mW the response of the material changes as the intensity of the $E_{2g}^1$ peak increases relative to that of $J_3$. At these powers the percentage of the 2$H$ phase is increasing exponentially. Above 1.5 mW the ratio of the two intensities from the flake with a thickness of 23 nm, slightly drops, which indicates degradation effects. It has been reported that damage and etching occur for flakes with similar thicknesses above 1 mW [32]. Another feature of this plot is that the data follow an "exposure-response" relationship, meaning that the phase transition of the MoS$_2$ shows a sigmoidal dependence on the radiation power, similar to the growth behavior of several materials such as graphene [33], silver nanoparticles [34] and nanowires [35]. From a fit to the data by a Boltzmann sigmoidal function and the extrapolation of the linear part we obtain the threshold power of about 0.5 mW, were the conversion of the 1$T$/1$T'$ to the 2$H$ phase takes place for the 6 nm thick sample. For the flake with a thickness of 23 nm we find a threshold of 0.4 mW. This slightly lower value is expected as a higher fraction of the incoming photons is absorbed by the thicker MoS$_2$. It is worth mentioning that while the $E_{2g}^1$ and $A_{1g}$ Raman peaks are restored, this does not occur for the exciton peaks in the photoluminescence spectrum, suggesting non-radiative recombination processes between the electron-hole pairs. Possible mechanisms are electron-hole separation due to remaining 1$T$/1$T'$ phase patches [36] or relaxation of the exited electrons through gap states from defects. This observation is in agreement with studies of the photoluminescence spectra from chemically exfoliated few layer-MoS$_2$ (~10nm) after annealing [37].

Moving a step further, by exploiting laser patterning, we study the creation of lateral heterostructures of 1$T$/1$T'$- and 2$H$-MoS$_2$ (Fig. 3a). The devices were fabricated from mechanically exfoliated MoS$_2$ flakes, transferred onto SiO$_2$ (285 nm)/Si substrates using the dry transfer technique explained in ref. [38], followed by standard electron beam (*e*-beam) lithography and metal deposition of Ti/Au contacts. After the devices were immersed in BuLi for more than 24 h for the transformation to take place. The flakes were subsequently exposed to laser radiation and the resulting electrical characteristics were studied.

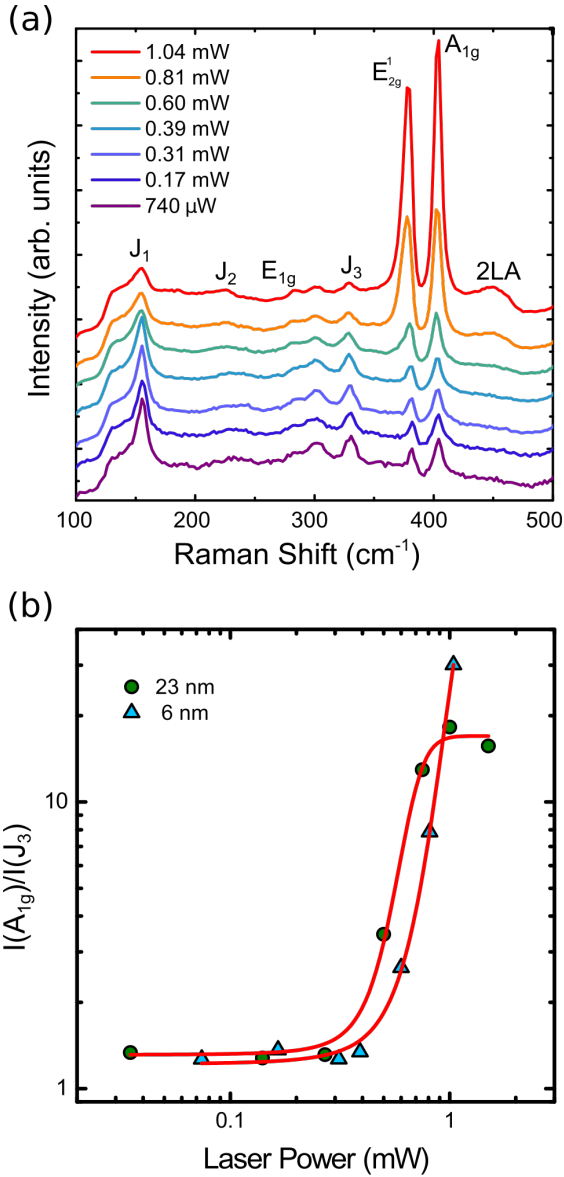

Fig. 2. Raman spectroscopy of laser induced annealing on the 1$T$/1$T'$-MoS$_2$. (a) Evolution of the Raman spectrum of a 6 nm 1$T$/1$T'$-MoS$_2$ flake with increasing the laser power. As the laser power increases the $E_{2g}^1$ and $A_{1g}$ peaks from the 2$H$ phase of the MoS$_2$ become more prominent. (b) Relative intensity ratio of the $A_{1g}$ to $J_3$ peak as a function of the laser power accompanied by Boltzmann sigmoidal function fit for two flakes with thickness 6 nm and 23 nm. The behavior of the two different in thickness flakes is similar.

Most of the lithiated $MoS_2$ devices show weak modulation of the conductance with the back-gate voltage (Fig. 3(c)). This has been observed in other studies and it is due to the high concentration of the carriers in the channel [27], [37], [39]. Furthermore, in one case (see device D in Fig. 3(c)) we observed a relatively stronger and positive transconductance, possibly as a result of a higher content of 2H phase remnants within the channel.

For the patterning of the semiconducting channel of the devices we scribed a ~0.5 μm long line across the width of the devices with a laser power of ~1 mW per spot that has a diameter of 500 μm. Figure 3(b) shows an atomic force microscopy (AFM) image of the topography of device with a laser patterned channel. The laser treated narrow channel between source and drain is evident. In agreement with Yinsheng Guo *et al.* [20], we find that laser treated areas have lower thickness than the untreated ones (7 nm for the untreated and 5.8 nm for the laser patterned part). This change in the thickness is around 20-30%. Chemically exfoliated samples of $1T/1T'$-$MoS_2$ are also thicker than flakes of mechanically exfoliated MoS$_2$ [37]. A possible explanation is the difference in the structure of the two phases as suggested in ref. [20]. Also, the possibility of intercalation of LiOH or other sources of contaminants between the layers from the chemical treatment, should not be excluded.

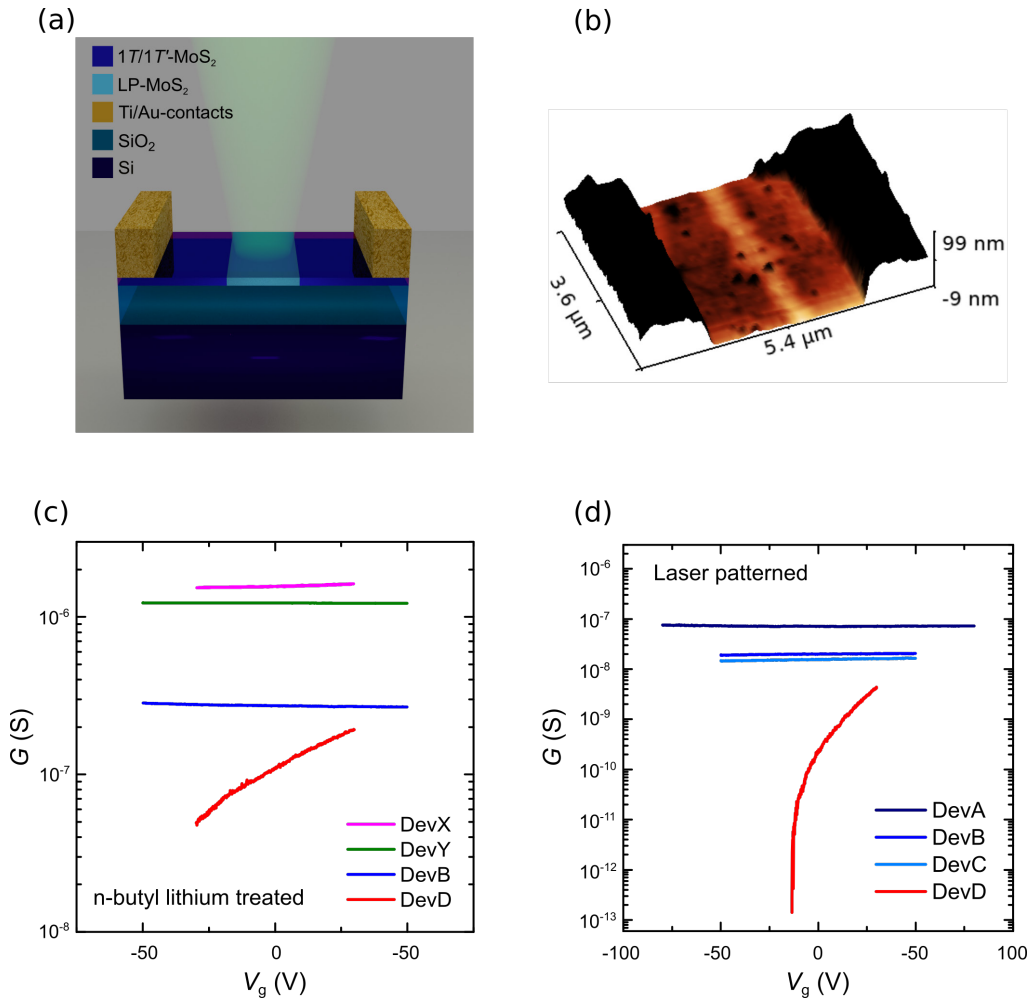

Fig. 3. Investigating the laser induced phase transitions of a few layer $MoS_2$ FETs. (a) Illustration of a device where the laser is patterning a strip of the flake to create the semiconducting 2H-$MoS_2$ channel. (b) 3D AFM topographic image of a device (B) after laser patterning of a narrow strip in the middle of the channel with a laser power of ~1 mW and spot size of 500 nm. The strip in the middle of the channel has an approximate width of 400 nm and corresponds well with the laser treated portion of the $MoS_2$. (c) Conductance as a function of the back-gate voltage from four $1T/1T'$-$MoS_2$ devices directly after the chemical treatment with BuLi. The poor modulation of the conductance from the back-gate voltage indicates the high carrier concentration in the devices. (d) Conductance as a function of the back-gate voltage from four devices after patterning a strip in the middle of the transistors as illustrated in (a) and shown in (b). In most devices the modulation of the conductance from the back-gate voltage is still negligible, which illustrates the ineffective laser induced phase change.

After laser exposure we investigated the electrical behavior of several devices and we did not observe a full restoration of the intrinsic semiconducting properties from the channel in most of the devices (Fig. 3(d)). The transconductance in most of the channels became positive but with a negligible ON/OFF ratio (see device A, B and C in Fig. 3(d)). Furthermore, a tenfold decrease in the conductance compare to the BuLi treated devices was observed that can be attributed to possible defect formation or oxidation from the exposure to the radiation.

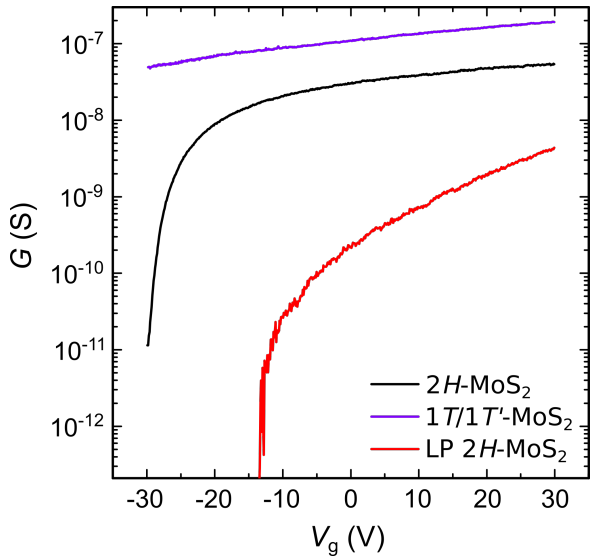

Fig. 4. Comparison of the different treatments to the characteristics of device D. Conductance as a function of the back-gate voltage for the intrinsic 2$H$-MoS$_2$ (black), after the treatment with BuLi (blue) and after the laser patterning (LP) of a semiconducting channel (red).

In one case, a restoration of the semiconducting properties of the device was found. To compare the device characteristics before any treatment and after the final laser patterning, we plotted the conductance of device D (with thickness of 3 layers), as a function of the gate voltage, for the intrinsic channel, after the treatment with BuLi and after the laser patterning of a thin semiconducting strip. As it can be seen, the conductance of the device has been reduced by more than an order of magnitude after the laser patterning, in comparison to the intrinsic channel. Moreover, the field-effect mobility that was calculated based on the relation: $\mu = \frac{L}{W}\frac{dG}{dV_g}\frac{1}{C_{ox}}$, with $L$ the length of the channel, $W$ the width of the channel and $C_{ox}$=1.21×10$^{-4}$ Fm$^{-2}$ the gate capacitance per unit area, reduced from 3.2 cm$^2$V$^{-1}$s$^{-1}$ before the chemical treatment to 1.1 cm$^2$V$^{-1}$s$^{-1}$ after the laser patterning. On the other hand, there was an improvement to the values of the sub-threshold swing which decreased to 0.6 V/dec.

The fact that even after the laser patterning there is no restoration of the semiconducting behavior in most of our devices, while the Raman spectra show the reappearance and increase in the intensity of the E$_{2g}$ and A$_{1g}$ peaks, is not surprising. Recent studies on the phase conversion of Li treated MoS$_2$ through X-ray photoelectron spectroscopy (XPS) and Raman showed similar discrepancies between the two techniques [40]. While the Raman peaks of the 1$T$ phase reduced and the peaks of the 2$H$ phase increase upon laser annealing, the XPS data show that there is still a significant amount of 1$T$/1$T'$ phase content within the flakes. Also FETs which were annealed at 123 °C, still show negligible resistance variation when the back-gate is applied [40]. This discrepancies among the different characterization techniques attributed to the different cross sections of the Raman scattering between the different MoS$_2$ phases. Therefore, the reason for the very weak transconductance and low on/off ratio in our laser patterned devices is most likely due to a significant remaining content of 1$T$/1$T'$-MoS$_2$, which screens the electric-field from the back-gate and is undetectable with the Raman spectroscopy. This could also be the explanation of the non-restored photoluminescence spectra as noted above. The difference between restored annealed devices from previous studies and the laser patterned ones here, is that in the former case annealing takes place in inert environment and for longer durations for a successful restoration [37],[24]. The laser induced local annealing under normal conditions here seems to be insufficient to restore completely the 2$H$ phase within 30 seconds and is most likely additionally promotes defect formation and even oxidation [40] which results in lower conductance of laser patterned devices. The degradation of the devices during exposure could be resolved if such laser scribing process would be performed under vacuum or inert atmosphere conditions.

### III. Conclusions

In summary, we have studied the metallic to semiconducting phase transition in thin MoS$_2$ flakes by radiation with a CW green laser. We found that the laser power range in which the intensity Raman peaks from the

2H phase start to increase, is in the range of 0.3 to 1 mW. Using controlled laser patterning in several MoS$_2$ devices, we find that in most of the devices the *n*-type semiconducting characteristic of MoS$_2$ are not restored, while the electrical properties degrade possibly due to defect formation or oxidation of the exposed area. This work shows that the polymer-based patterning and chemical treatment [23], is the best route to obtain on demand novel heterostructures from the various MoS$_2$ phases. Lastly our work suggests that any assessment of metallic to semiconducting transitions in TMDCs should be realized via photoluminescence or XPS characterization techniques.

ACKNOWLEDGMENTS

We would like to thank Damien Voiry for his suggestions regarding the chemical intercalation process, Wiel Ewers for his assist at the glove box facilities and Andres Castellanos-Gomez for measurements at his facilities and for discussions.